# A revisit of superconductivity in $4H_b$-TaS$_{2-2x}$Se$_{2x}$ single crystals

Zhenkai Xie, Meng Yang, Zhigang Cheng, Tianping Ying, Jian-Gang Guo, Xiaolong Chen

Beijing National Laboratory for Condensed Matter Physics, Institute of Physics, Chinese, Academy of Sciences, Beijing, 100190, China
University of Chinese Academy of Sciences, Beijing, 100049, China.
Songshan Lake Materials Laboratory, Dongguan, Guangdong 523808, China.

**Abstract**
Previous investigations of $4H_b$-TaS$_{2-2x}$Se$_{2x}$ mainly focused on the direct competition between superconductivity and charge density wave (CDW). However, the superconductivity itself, although has been prominently enhanced by isovalent Se substitution, has not been adequately investigated. Here, we performed a detailed electrical transport measurement down to 0.1 K on a series of $4H_b$-TaS$_{2-2x}$Se$_{2x}$ single crystals. A systematic fitting of the temperature-dependent resistance demonstrates that the decreased Debye temperatures ($\Theta_D$) and higher electron-phonon coupling constants ($\lambda_{e-p}$) at the optimal Se doping content raise the superconducting transition temperature ($T_c$). Additionally, we discovered that the incorporation of Se diminishes the degree of anisotropy of the superconductivity in the highly layered structure. More prominently, a comprehensive analysis of the vortex liquid phase region reveals that the optimally doped sample deviates from the canonical 2D Tinkham prediction but favors a linear trend with the variation of the external magnetic field. These findings emphasize the importance of interlayer interaction in this segregated superconducting-Mott-insulating system.

**Introduction**
    TaS$_2$ is one of the most investigated transition metal chalcogenides with rich crystal structures and physical properties [1][2][3][4]. Depending on its synthesis condition, the building block of TaS$_6$ will adopt either octahedral coordination (T phase) at high temperatures or trigonal prismatic coordination (H phase) at low temperatures with distinct electrical transport behaviors and ground states. The majority of the research has focused on the interaction between superconductivity and CDW state. For example, the 1T-TaS$_2$ is known as a Mott insulator with three successive phases of charge density wave from incommensurate (550 K) to nearly commensurate (350 K) to commensurate (180 K) by decreasing temperature [1][6]. The 1T phase can be driven into a superconductor by substitution, intercalation, distortion, or application of external pressure [6][7][8][9]. On the other hand, 2H-TaS$_2$ shows a metallic behavior with the coexisting of a commensurate CDW at 78 K and superconductivity at 0.8 K [10]. Isovalent Se substitution of the 2H-TaS$_{2-2x}$Se$_{2x}$ generates a superconducting plateau with a similar $T_c$ of ~3.3 K in a wide doping range of 80% (0.1<x<0.9) [11].

    Using the van der Waals nature of the transition metal chalcogenides, TaS$_2$ can be

spontaneously assembled into a superlattice with alternating T and H phases at a moderate growth temperature, known as the $4H_b$ phase, with coexisting superconductivity at 2.7 K and CDW ordering at 22 K and 315 K [3][12][13]. Different from $2H$-$TaS_2$ where superconductivity and CDW take place at the same layer, it is suggested that the T-layer in $4H_b$-$TaS_2$ is solely responsible for the $\sqrt{13} \times \sqrt{13}$ commensurate CDW state forming a periodic Star-of David clusters below 315 K. Meanwhile, the H-layer contributes to the appearance of superconductivity and a weak 3×3 CDW superlattice below 22 K [14][15][16] [17]. The significant $T_c$ enhancement from 0.8 to 2.7 K and the effective CDW suppression from 78 to 22 K are ascribed to the dimensional reduction from 2H phase into 1H layer. Nevertheless, the primary feature of superconductivity followed by a 3×3 charge modulation in the 1H layer resembles that of the 2H phase [18]. Following this dichotomy interpretation, we would expect to see a similar wide $T_c$ plateau in isovalent Se doping as that of $2H$-$TaS_{2-2x}Se_{2x}$ [10]. However, $4H_b$-$TaS_{2-2x}Se_{2x}$ shows a volcano-like superconducting phase diagram with the maximum $T_c$ appearing at x~30% [19]. This observation implies that the interlayer interaction from the Mott-insulating 1T-layer should also pose an influence on its adjacent 1H layer, but the underlying mechanism is not clear. Moreover, to the best of our knowledge, only the out-of-plane superconducting properties that have been reported, while the in-plane information remain unexplored. It is also not clear whether the 2D nature of the superconductivity, generally described in the framework of Tinkham theory, will be preserved, strengthened, or diminished. It is thus imperative to carry out a systematic investigation of the $4H_b$-$TaS_{2-2x}Se_{2x}$ for a comprehensive understanding.

In this paper, we measure the electrical transport properties of a series of $4H_b$-$TaS_{2-2x}Se_{2x}$ single crystals down to 0.1 K along different crystallographic directions. By fitting the Bloch-Grüneisen model for each $\rho$-$T$ curve, we successfully extracted the Debye temperature ($\Theta_D$) and electroacoustic coupling constant ($\lambda_{e-p}$). We found that $T_c$ is negatively correlated with $\Theta_D$ and positively correlated with $\lambda_{e-p}$, which ascribes the optimal $T_c$ to the emergence of certain vibrational modes. The study of the in-plane upper critical field allows us to compare the changes in the anisotropy of the system upon isovalent Se doping, which reveals an interesting isotropic tendency and a clear deviation to the canonical 2D Tinkham formula. Our results suggest the isovalent intralayer substitution could introduce interlayer interactions in the $4H_b$ system.

**Experimental**
The series of single crystals $4H_b$-$TaS_{2-2x}Se_{2x}$ are synthesized through the conventional chemical vapor transport (CVT) method using iodine as the transport agent [12]. The sealed quartz tubes were placed in a two-zone furnace for one week with the hot and cold ends kept at 930 °C and 830 °C, respectively. After the crystal growth, the samples are harvested by quenching the tube in water to preserve the $4H_b$ phase. The powder X-ray diffraction (XRD) patterns were taken using a Bruker D8 Advance diffractometer with Cu-Kα radiation at room temperature. Low-temperature transport (down to 100 mK) measurements were performed using a dilution refrigerator. Transport, magnetic and thermodynamic properties were measured using the Physical Property Measurement System (PPMS, Quantum Design) and SQUID vibrating sample magnetometer (SVSM, Quantum Design). Elemental compositions were determined using an electron-probe microanalyzer. The compositions were probed at around ten focal points, and the results were averaged.

**Results and Discussion**

The crystal structure of $4H_b$-TaS$_{2-2x}$Se$_{2x}$ is shown in Fig. 1a by stacking 1T and 1H-layer along the *c*-axis. A series of high-quality single crystals can be feasibly grown into centimeter size by the chemical transport growth method as described in the Experimental section. The whole diffraction patterns can be found in Supplementary Information Fig. S1 without appreciable impurities. The enlargement of the (004) diffraction peaks is enlarged in Fig. 1b, from which the systematic peak shift to lower angles can be observed, indicating the increase of the *c* axis as shown in Fig. 1c. This lattice expansion is consistent well with the previous report [19].

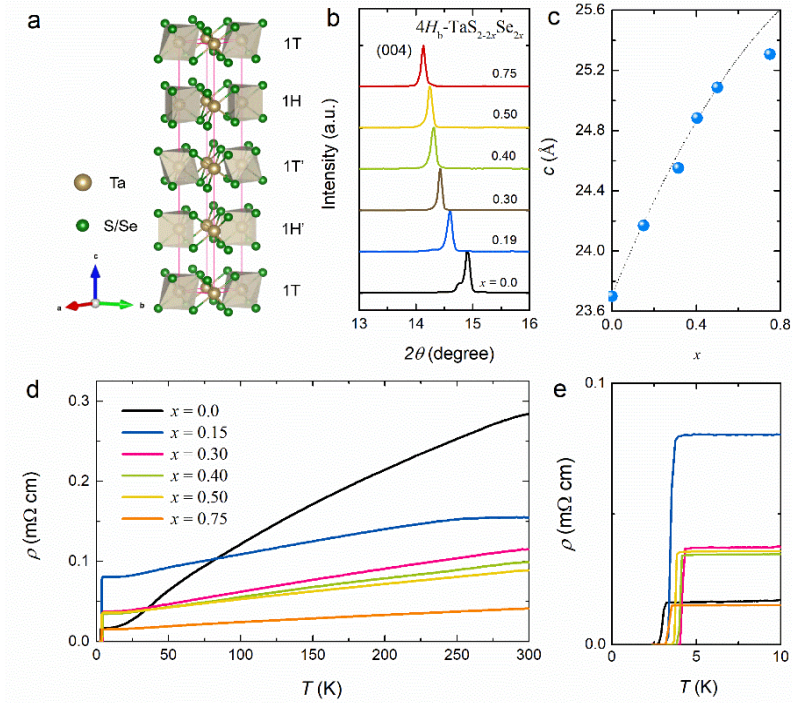

**Figure 1** (a) Crystal structure of $4H_b$-TaS$_{2–2x}$Se$_{2x}$ (x=0.0-0.75). (b) The evolution of diffraction peak along the (004) direction of $4H_b$-TaS$_{2–2x}$Se$_{2x}$; (c) The indexed lattice constant *c* of $4H_b$-TaS$_{2–2x}$Se$_{2x}$ (x=0.0-0.75). (d) Temperature dependence of in-plane resistivity *ρ* of $4H_b$-TaS$_{2–2x}$Se$_{2x}$ below 300 K with the enlargement of its superconducting region shown in (e).

As the general competition trend of CDW ordering and the superconductivity has already been established in Ref. 19, we concentrate on the metallic behavior below the CDW order as well as the superconducting region. The temperature-dependent in-plane resistivity for various doping compositions is shown in Fig. 1d, with its enlargements shown in Fig. 1e. A prominent feature is the dramatic change of the slope in the normal state from the raw $4H_b$-$TaS_2$ to its Se-doped counterparts. Therefore, we fit each curve by the Bloch-Grüneisen function (shown in Fig. S2)

$$\rho(T) = \rho_0 + \frac{4B}{\Theta_D}\left(\frac{T}{\Theta_D}\right)^n \int_0^{\Theta_D/T} \frac{z^n dz}{(e^z-1)(1-e^{-z})} - kT^3$$

$\rho_0$ is residual resistivity, $\Theta_D$ Debye temperature, $B$ electron-phonon coupling coefficient [20][21][22] and plot these values in Fig. 2. The Bloch-Grüneisen function describes the data up to room temperature, even above $\Theta_D/4$, indicating significant s-d band scattering. While the value of $n$ is typically fixed to be 2, 3, or 5. [23]

An interesting discovery is the extracted Debye temperature roughly inversely evolves with the $T_c$ at x=0.3 (Fig. 2b), indicating the phonon component should be the culprit for the volcano-like $T_c$ phase diagram. Thus, we fit the Mcmillan formula [24] using the extracted $T_c$ and $\Theta_D$:

$$\lambda_{e\text{-}p} = \frac{1.04 + \mu^* \ln\left(\frac{\Theta_D}{1.45 T_C}\right)}{(1-0.62\mu^*)\ln\left(\frac{\Theta_D}{1.45 T_C}\right)-1.04},$$

where the variable $\mu^*$ represents the repulsive screened Coulomb potential with typical material-specific values in the range $0.1 \leq \mu^* \leq 0.15$. Typically, materials with $\lambda_{e\text{-}p} \to 1$ are classified as strongly coupled superconductors, while $\lambda_{e\text{-}p} \to 0.5$ indicates weak coupling [23] Here, the Coulomb pseudo-potential $\mu^*$ is taken as 0.15. The calculated $\lambda_{e\text{-}p}$ shown in Fig. 2d reproduces the volcano shape with a maximum at x = 0.3. Within the framework of BCS theory, the negatively correlated Debye temperature and the positively correlated lambda with the $T_c$ diagram imply the $T_c$ enhancement in the $4H_b$-$TaS_{2-2x}Se_{2x}$ is driven by the involvement of softening phonons [25].

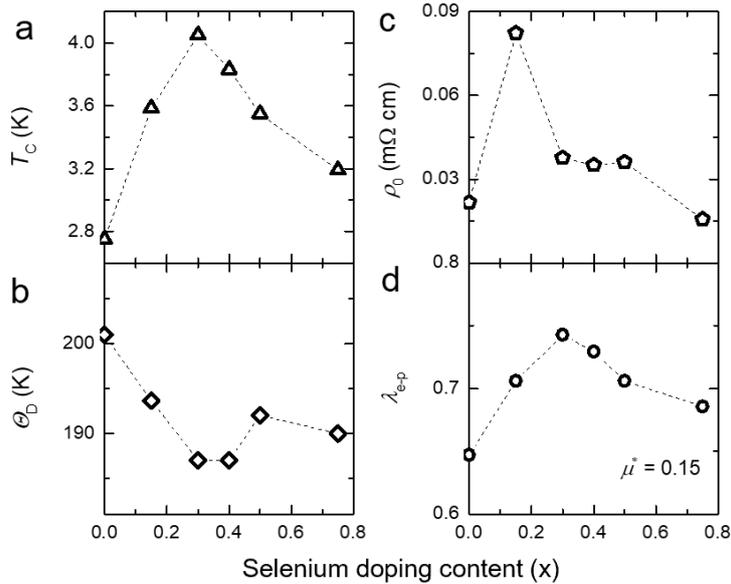

**Figure 2** Evolution of (a) superconducting temperature $T_c$, (b) Debye temperature $\Theta_D$ (c) residual resistivity $\rho_0$, and (d) electron-phonon constant $\lambda_{e\text{-}p}$ of $4H_b$-$TaS_{2-2x}Se_{2x}$

(x=0.0-0.75), obtained by the fitting of $\rho$(T) data based on Bloch-Grüneisen model [23].

Previous research on the $4H_b$-TaS$_{2-2x}$Se$_{2x}$ single crystals mainly focused on out-of-plane information. Therefore, we investigate the transport properties with the external magnetic field applied with the ab plane, from which we could analyze the anisotropy of its superconducting properties. Three typical samples with the magnetic field applied both perpendicular and within the ab plane are illustrated in Fig. 3. We define the drop of the resistance to 50% of their normal state value as its $T_c$ for comparison with Ref. 3. We summarized these $H_{c2}$ in Fig. 4 together with the Ginzburg-Landau fitting curves shown in broken lines [26][27]. The data points of the pristine material $4H_b$-TaS$_2$ are extracted from Ref. 3. As shown in Figs 4a and 4b, the $H_{c2}^{ab}$ remains almost identical at around 17 T, the $H_{c2}^{c}$ is dramatically enhanced from 1 T to 2-3 T, which means the diminish of the anisotropic superconductivity. To quantify this change, we define $\gamma = \frac{H_{c2}^{ab}}{H_{c2}^{c}}$ as the anisotropy of the superconductivity. A dramatic decrease from $\gamma$ = 17 in $4H_b$-TaS$_2$ to $\gamma$ = 7 in the Se-doped samples can be clearly seen in Fig. 4c, implying the superconductivity becomes more isotropic upon isovalent Se substitution.

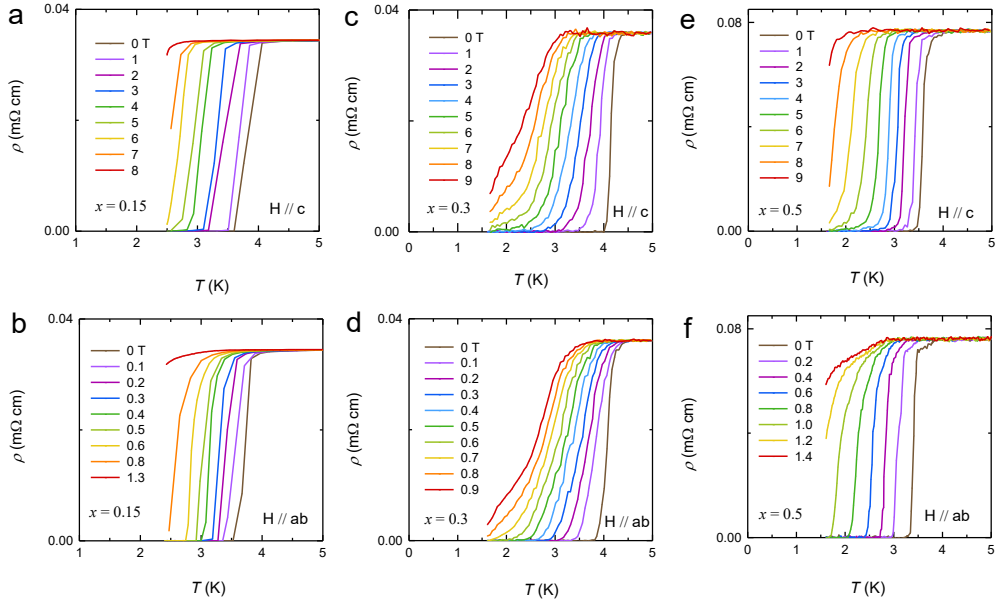

**Figure 3** Temperature-dependent resistivity $\rho$ of $4H_b$-TaS$_{2-2x}$Se$_{2x}$ (x=0.15, 0.3, and 0.5) under a series of constant magnetic fields parallel to *c* axis (a, c, and e) and parallel to *ab* plane (b, d, and f).

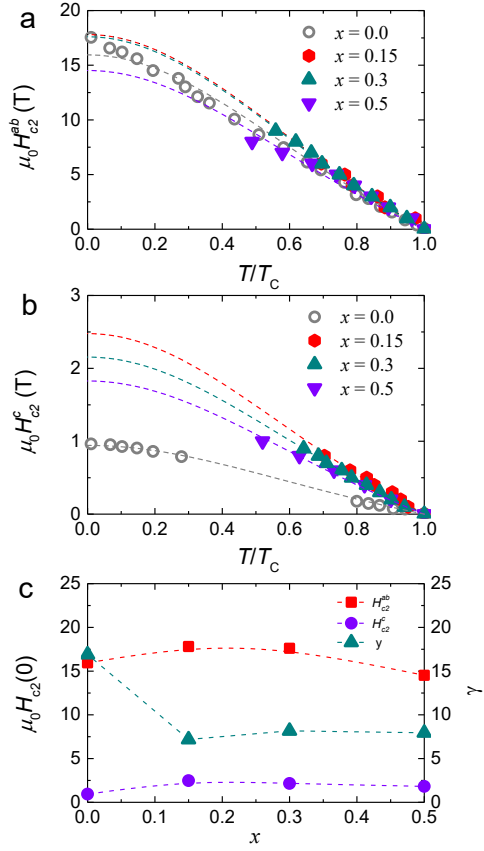

**Figure 4** Temperature-dependent in-plane (a) and out-of-plane (b) upper critical field of $4Hb$-TaS$_{2-2x}$Se$_{2x}$ ($x$ = 0.0, 0.15, 0.3, 0.5). Noted that the data (marked as gray hollow circle) of $4H_b$-TaS$_2$ is extracted from Ref. 3. The dash line is the fitting results of the Ginzburg-Landau model [26]; (c) Evolution of in-plane upper critical field $H_{c2}^{ab}$, out-of-plane upper critical filed $H_{c2}^{c}$ and anisotropy factor $\gamma$ of $4H$-TaS$_{2-2x}$Se$_{2x}$ ($x$ = 0.0, 0.15, 0.3, 0.5).

To a certain extent, the sandwiching 1H layer (superconducting) and 1T (Mott insulating) heterostructure resembles the recently discovered clean 2D superconductor Ba$_6$Nb$_{11}$S$_{28}$ [28] with superconducting 2H-NbS$_2$ separated by insulating Ba$_3$NbS$_5$ layer. This is indeed the case that the anisotropy of $\gamma$ = 4.5 in 2H-TaS$_2$ [29] can be dramatically enhanced to $\gamma$ = 17 in $4H_b$-TaS$_2$. But it is not easy to understand why the isovalent Se doping could effectively introduce interlayer interactions. We therefore resort to the analysis of the vortex state in the optimal doping. The upper critical field $H_{c2}$ could give a deeper insight into the nature of the superconductivity [30][31][32]. To ensure the correctness of our analysis with the sufficient fitting region, the field-dependent resistance is measured down to 0.1 K as shown in Fig. 5. The 90% of the normal resistance is taken as the criteria of $H_{c2}$ for each temperature, and the critical points of the resistance deviate from zero are taken as the irreversible magnetic field $H_{irr}$ [29][33]. These data points are summarized in Figs 5b and 5d, where the direction-dependent $H$-$T$ phase diagram can be divided into three regions including vortex solid, vortex liquid, and normal states. The horizontal temperature difference between $H_{c2}$ and $H_{irr}$ for a given external field are defined as $\Delta T$. According to Tinkham's theoretical prediction

[34], canonical 2D superconductors will obey a scaling rule of $\Delta T$ vs. $\mu_0 H^{2/3}$. This is generally the case for many layered superconductors including $YBa_2Cu_3O_{7-\delta}$ [35], 2H-$NbSe_2$[36], and 2H-$TaS_2$ [29]. However, a clear deviation from the $H^{2/3}$ rule is shown in Fig. 5e, indicating the Tinkham model no longer holds in the $4H_b$-$Ta_{2-2x}Se_{2x}$ system. Interestingly, we find a simple linear fitting of $\Delta T \sim H$ agrees with our data better. The underlying mechanism is not clear and is worthy of further investigations.

This breakdown of the 2D Tinkham formula can also be reflected in the analysis of the coherence length (Fig. S3). We choose $H_{c2,50\%}$ in Fig.5a and Fig.5c for our analysis to include more data points. Different choices will not change the conclusion. The $H_{c2}^c$ and $H_{c2}^{ab}$ can be best described using the GL function to be 3.38 T and 18.73 T, respectively. Their coherence lengths can be feasibly estimated using the formula

$$H_{c2}^c(0) = \frac{\Phi_0}{2\pi\xi_{ab}^2}; \quad H_{c2}^{ab}(0) = \frac{\Phi_0}{2\pi\xi_{ab}\xi_c}$$

which gives the in-plane (out-of-plane) Ginzburg-Landau coherence length $\xi_{ab}$ ($\xi_c$) to be 99 Å (18 Å), indicating the interlayer interaction cannot be neglected. Fitting of the in-plane (out-of-plane) GL coherence length in 4Hb-$TaS_2$ gives 186 Å (9.8 Å) [3]. It should be noted that both the low-temperature $H_{c2}^c$-T and $H_{c2}^{ab}$-T curves exceed the single-band Werthamer-Helfand-Hohenberg model [37], which may be attributed to multiband effects [38]. Therefore, we suggest that the isovalent Se doping into $4H_b$-$TaS_{2-2x}Se_{2x}$ can not be viewed as a simple superposition of two independent Se-substitution processes into the 1T and 1H layers. One plausible explanation is the Se substitution into 1T-$TaS_{2-x}Se_x$ could effectively increase the conductivity of the 1T phase and eventually induce superconductivity [39][40]. Without the segregation of insulating 1T layer, the interlayer hopping between the 1H layer through 1H-1T-1H will be enhanced, and finally, reduce the anisotropy properties of the $4H_b$ system. Such kind of interlayer interaction may also explain the appearance of a volcano-like superconducting phase diagram, where the interlayer attraction force may reduce the specific vibrational modes.

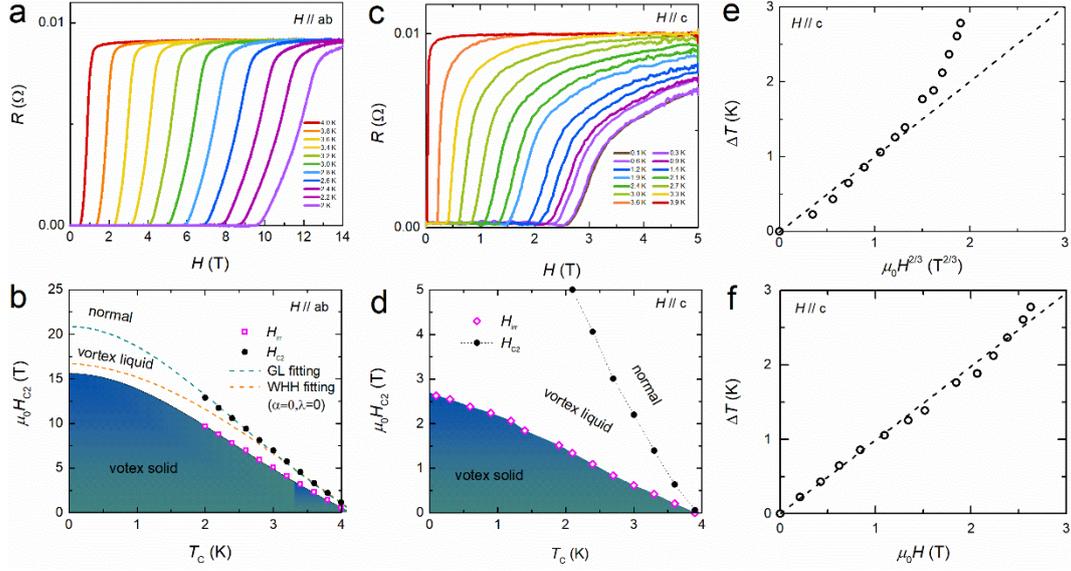

**Figure 5** Field-dependent resistivity $\rho$ of $4H_b$-$TaS_{1.4}Se_{0.6}$ under a constant in-plane (a) and out-of-plane (c) magnetic field down to 0.1 K. The temperature-dependent upper critical field $\mu_0 H_{c2}$ and the irreversible field, $\mu_0 H_{irr}$ for $H\|ab$ (b) and $H\|c$ (d). Black circles represent the upper critical field $H_{c2}(T)$ data. Magenta symbols represent the irreversibility field $H_{irr}(T)$ obtained from the field-dependent resistivity data. The orange and green dashed lines show GL fitting and WHH fitting results of $H_{c2}(T)$, respectively. The $H_{c2}$ is obtained by using the criterion of $90\%\rho_n(H)$ in $R$–$H$ curves, and the irreversible magnetic field $H_{irr}$ was obtained from the zero value of $T_c$ in in $R$–$H$ curves. The width of vortex liquid region is scaled as (e) $\Delta T$ vs. $\mu_0 H^{2/3}$ and (f) $\Delta T$ vs. $\mu_0 H$.

**Conclusion**

In this paper, we focus on the anisotropy properties of the $4H_b$-$TaS_{2-2x}Se_{2x}$ system by measuring its resistivity down to 0.1 K. Several new features are revealed including the enhanced $\lambda_{e-p}$ at $T_c^{max}$, decreased superconducting anisotropy and breakdown of the Tinkham model. The inclusion of isovalent Se greatly alters the electrical property in the 1T layer which in turn dramatically enhances the interlayer interaction with its adjacent 1H layers. Such a hybridized material system thus provides a natural platform for the investigation of the intricate interplay among low-dimensional superconductivity, CDW states and Mott physics.

**Acknowledgments**

This work is financially supported by the National Key Research and Development Program of China (nos. 2017YFA0304700, and 2021YFA1401800), Beijing Natural Science Foundation (grant no. Z200005).